\documentclass[]{llncs2e/llncs}

\usepackage{graphicx}
\usepackage{algorithm}
\usepackage{algorithmicx}
\usepackage[noend]{algpseudocode}
\usepackage[disable]{todonotes}
\usepackage{amssymb}
\usepackage{amsmath}
\usepackage{verbatim}

\title{The Sharer's Dilemma in Collective Adaptive Systems of Self-Interested Agents}
\author{Lenz Belzner\inst{1} \and Kyrill Schmid\inst{2} \and Thomy Phan\inst{2} \and Thomas Gabor\inst{2} \and Martin Wirsing\inst{2}}
\institute{MaibornWolff \and LMU Munich}

\begin{document}

\maketitle

\begin{abstract}
In collective adaptive systems (CAS), adaptation can be implemented by optimization wrt. utility. Agents in a CAS may be self-interested, while their utilities may depend on other agents' choices. Independent optimization of agent utilities may yield poor individual and global reward due to locally interfering individual preferences. Joint optimization may scale poorly, and is impossible if agents cannot expose their preferences due to privacy or security issues.

In this paper, we study utility sharing for mitigating this issue. Sharing utility with others may incentivize individuals to consider choices that are locally suboptimal but increase global reward. We illustrate our approach with a utility sharing variant of distributed cross entropy optimization. Empirical results show that utility sharing increases expected individual and global payoff in comparison to optimization without utility sharing.

We also investigate the effect of greedy defectors in a CAS of sharing, self-interested agents. We observe that defection increases the mean expected individual payoff at the expense of sharing individuals' payoff. We empirically show that the choice between defection and sharing yields a fundamental dilemma for self-interested agents in a CAS.
\end{abstract}

\section{Introduction}

In collective adaptive systems (CAS), adaptation can be implemented by optimization wrt. utility, e.g. using multi-agent reinforcement learning or distributed statistical planning \cite{tan1993multi,hillston2015collective,belzner2016collective,foerster2017stabilising,phan2018evade}. Agents in a CAS may be self-interested, while their utilities may depend on other agents' choices. This kind of situation arises frequently when agents are competing for scarce resources. Independent optimization of each agent's utility may yield poor individual and global payoff due to locally interfering individual preferences in the course of optimization \cite{leibo2017multi,perolat2017multi}. Joint optimization may scale poorly\todo{reference}, and is impossible if agents do not want to expose their preferences due to privacy or security issues \cite{brundage2018malicious}.

\begin{figure}
	\centering
	\includegraphics[scale = 0.4]{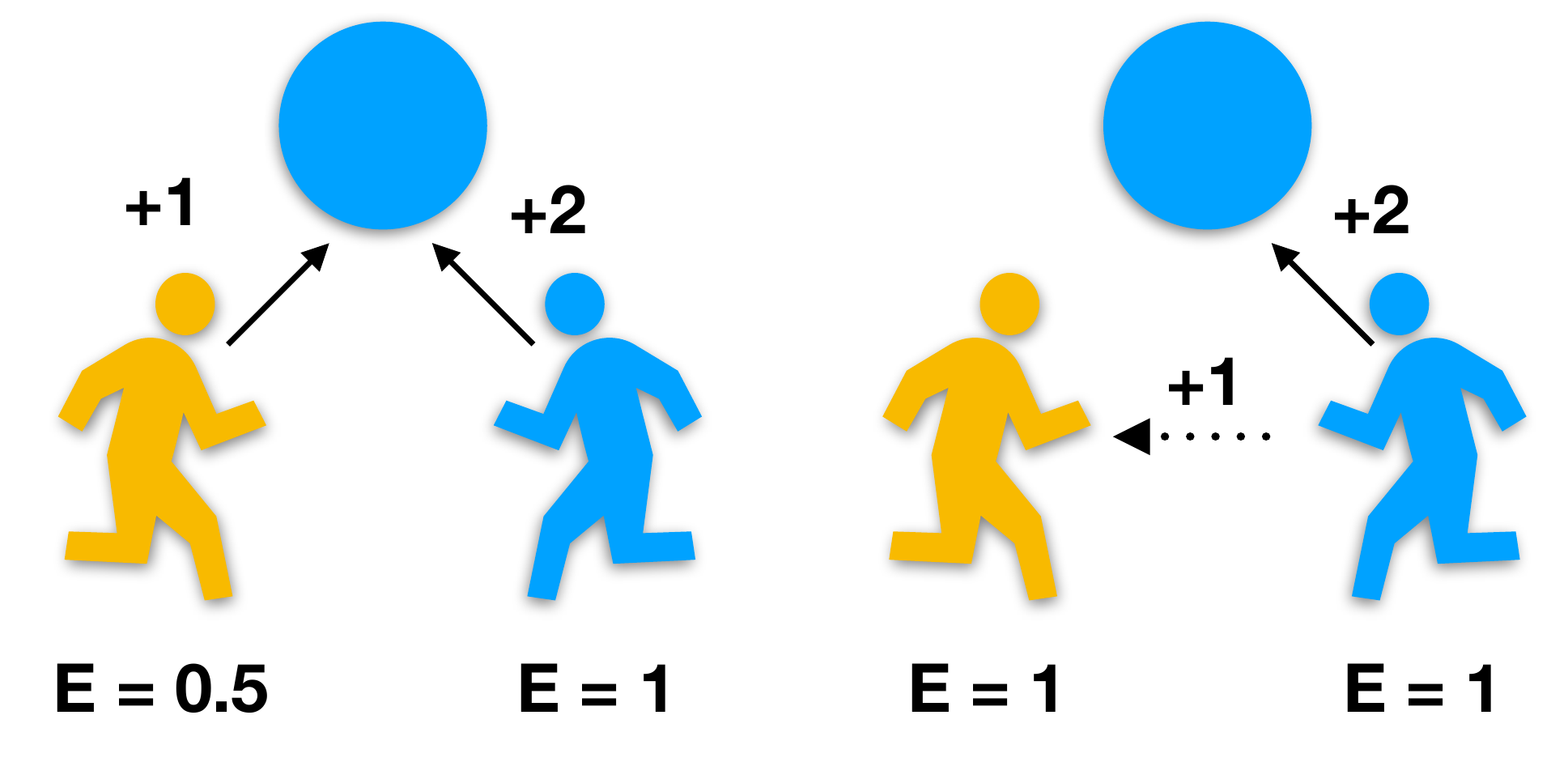}
	\caption{Two agents competing for a coin: if agent 1 (yellow) on the left side happens to get the coin it will get a reward of +1 whereas agent 2 (blue) will get a reward of +2 for it. If there is a fifty-fifty chance for an agent to get the coin when both agents are trying to collect it, the expected values are 0.5 for agent 1 and 1 for agent 2 when both agents independently optimize their utility. In contrast, if there is the possibility to share reward then agents could learn to do the following: agent 1 (yellow) resists to collect the coin. That increases the blue agent's probability for getting a reward to 1. The blue agent transfers reward (e.g. 1) to the yellow agent. This leaves agents with expected values of 1 each and therefore defines a strong Pareto improvement compared to the former outcome.
	}
	\label{fig:coin_game}
\end{figure}

A minimal example of such a situation is the coin game \cite{lerer2017maintaining} (cf. Figure \ref{fig:coin_game}. \todo{cite L2T} Here, a yellow and a blue agent compete for coins. The coins are also colored in yellow or blue. Both agents can decide whether to pick up the coin or not. If both agents opt to pick up the coin, one of them receives it uniformly at random. If an agent picks up a coin of its own color, it receives a reward of 2. If it picks up a differently colored coin, it gets a reward of one. Each agent wants to maximize its individual reward. If agents act purely self-interested, then each agent tries to pick up each coin, resulting in suboptimal global reward. However, if rewards can be shared among agents, then agents will only pick up coins of their own color. They receive a share that is high enough to compensate for not picking up differently colored coins. This increases individual and global reward alike.

There are many examples for this kind of situation. For example, energy production in the smart grid can be modeled in terms of a CAS of self-interested agents. Each participant has to decide locally how much energy to produce. Each agent wants to maximize its individual payoff by selling energy to consumers in the grid. However, the price is depending on global production. Also, global overproduction is penalized. Routing of vehicles poses similar problems. Each vehicle wants to reach its destination in a minimal amount of time. However, roads are a constrained resource, and for a globally optimal solution, only a fraction of vehicles should opt for the shortest route. In both scenarios, the ability of agents to share payoff may increase individual and global reward alike.

%


\todo {asymmetry \cite{tuyls2018symmetric}}
\todo{chicken/stag hunt, nash vs. optima, utility privacy as a motivation}


In this paper, we study distributed optimization with utility sharing for mitigating the issue of contrasting individual goals at the cost of expected individual and global reward. To illustrate our ideas, we propose a utility sharing variant of distributed cross entropy optimization. Empirical results show that utility sharing increases expected individual and global payoff in comparison to optimization without utility sharing.

We then investigate the effect of defectors participating in a CAS of sharing, self-interested agents. We observe that defection increases the mean expected individual payoff at the expense of sharing individuals' payoff. We empirically show that the choice between defection and sharing yields a fundamental dilemma for self-interested agents in a CAS.

The paper makes the following contributions.
\begin{itemize}
	\item We motivate utility sharing as a means to mitigate conflicts and increase expected individual and global reward in CAS of self-interested agents.
	\item We propose distributed optimization with sharing (DOS) as an algorithm to realize utility sharing in self-interested CAS.
	\item We evaluate DOS empirically, showing that it increases individual and global reward in expectation.
	\item We investigate the effect of defecting, non-sharing individuals in a group of self-interested sharing agents. We show that the choice between defection and cooperation yields a fundamental dilemma for self-interested agents in collective adaptive systems.
\end{itemize}

The remainder of the paper is structured as follows. In Section \ref{sec:related} we discuss related work. We introduce DOS in Section \ref{sec:dos}. We discuss our empirical results and the Sharer's Dilemma in Section \ref{sec:results}. We conclude in Section \ref{sec:conclusion}.

\section{Related Work}
\label{sec:related}

In general, we see our work in the context of collective adaptive systems (CAS) \cite{hillston2015collective,belzner2016collective} and multi-agent systems \cite{van2008multi}.
In particular, we are interested in CAS where agents are adaptive through optimization of actions of policies wrt. a given individual or global utility function. These settings can for example be modeled in terms of distributed constrained optimization problems \cite{fioretto2016distributed}, or as stochastic games \cite{shapley1953stochastic}.

Searching for optimal actions or learning policies can be done by open- or closed-loop planning, potentially enhanced with learned components such as search exploration policies or value functions \cite{phan2018evade,silver2016mastering,silver2017mastering,silver2017nature,anthony2017thinking}. Another approach for learning optimal policies in multi agent domains such as CAS is multi agent reinforcement learning (MARL) \cite{tan1993multi,littman1994markov} and its modern variants based on deep learning for scaling up to more complex domains\cite{foerster2017stabilising,foerster2016learning,tampuu2017multiagent}.
%
A recent example of planning-based deep MARL combines open-loop search and learned value functions in fully cooperative multi-agent domains \cite{phan2018evade}.

In the case of self-interested agents, the Coco-Q algorithm was proposed \cite{sodomka2013coco}. Coco-Q has been evaluated for discrete two-player matrix games, and requires explicit knowledge of other agents' utilities. In some sense, our study of sharing in CAS extends the Coco-Q approach to continuous optimization with more than two agents. Also, we model the amount sharing as a free parameter to be learned in the course of optimization.

In the context of a research on emergent social effects in MARL \cite{leibo2017multi,perolat2017multi,peysakhovich2017prosocial,lerer2017maintaining}, a recent report investigated the effects of inequity aversion and utility sharing in temporally extended dilemmas \cite{hughes2018inequity}. The authors state that "it remains to be seen whether emergent inequity-aversion can be obtained by evolving reinforcement learning agents" \cite{hughes2018inequity}. Our current work is a first step into this direction, and shows that the question of whether to share or not poses a dilemma in and for itself, at least in the case of stateless optimization (in contrast to learning policies).

\todo[inline]{game theoretic dilemmas (chicken, stag hunt, etc.), transferable utility, game theoretic concepts, explicit coordination mechanisms (in contrast to sharing and coordination emergence), evolutionary game theory, survival of fittest vs. survival of the tribe}

\section{Distributed Optimization with Sharing}
\label{sec:dos}

We model decision making in a CAS as a stochastic game $(X, N, A, p, R)$ \cite{shapley1953stochastic}.
\begin{itemize}
	\item $X$ is a finite set of states.
	\item $N = \{0, ..., n\}$ is a finite set of agents.
	\item $A = \times_{i \in N} A_i$ is a set of joint actions. $A_i$ is a finite set of actions for agent $i$.
	\item $p(x' \vert x, a)$ is a distribution modeling the probability that executing action $a \in A$ in state $x \in X$ yields state $x' \in X$.
	\item $R = \{r_i\}_{i \in N}, r_i : X \times A \rightarrow \mathbb{R}$ is a set of reward functions, one for each agent.
\end{itemize}

In the following, we assume $X = \{x\}$ consists of a single state, and $\forall a \in A : p(x \vert x, a) = 1$. As $x$ is unique, we will not consider it in further notation.

We assume that $r_i$ is available to agent $i$ in terms of a generative model that may be queried for samples $a$, e.g. a simulation of the application domain. Each agent only has access to its own reward function, but does not know the reward functions of other agents.

The task of a self-interested agent $i$ is to find an action that maximizes its payoff. However, its payoff $r_i(a), a \in A$ in general depends on the choices of other agents. One way to deal with this dependency is to perform optimization jointly for all agents, that is $\max_{a \in A} : \sum_{i \in N} r_i(a)$. However, in a CAS with self-interested agents, each participant tries to maximize its individual reward. Also, in many situations participating agents would not want to expose their individual reward functions to others due to privacy or security issues \cite{brundage2018malicious}. In these situations, joint optimization wrt. global reward is not feasible. Note that optimization of self-interested individuals is non-stationary due to changes in others' choices as they optimize for themselves.


\subsection{Reward Sharing}

We define agents' utilities as $u_i$. We consider the two different cases we are interested in:
\begin{enumerate}
	\item Individual, purely self-interested optimization
	\item Self-interested optimization with the option to share individual rewards
\end{enumerate}

\subsubsection{Pure Self-Interest}
When optimizing independently and purely self-interested, $u_i(a) = r_i(a)$.

\subsubsection{Sharing}
Sharing agents choose a share $s_i \in \mathbb{R}, s_i \geq 0$ additionally to $a_i$.
We denote the joint shares by $s = \times_{i \in N} s_i$.
Given $n$ agents, a joint action $a \in A$ and a joint share $s \in \mathbb{R}^n, s_i \geq 0$ for all $i$, we define individual agents' utility $u_i$ for distributed optimization with sharing as follows.
\footnote{We can account for the change of signature of $u_i$ by extending the action space $A_i$ of each agent accordingly: $A_{s, i} = A_i \times \mathbb{R}, A_s = \times_{i \in N} A_{s, i}$.}
\begin{equation}
\label{eq:dos_util}
u_i(a, s) = r_i(a) - s_i + \frac{\sum_{j, j \neq i} s_j}{n - 1}
\end{equation}


Shares are uniformly distributed among all other agents. There are no bilateral shares. Note that this sharing mechanism is an arbitrary choice.

For example, sharing yields the following utilities for two agents.
\begin{align*}
u_0(a, s) = r_0(a) - s_0 + s_1\\
u_1(a, s) = r_1(a) - s_1 + s_0
\end{align*}

%
%

\subsection{Distributed Optimization with Sharing}

We now give a general formulation of distributed optimization with sharing (DOS). DOS is shown in Algorithm \ref{alg:dos}.
Each agent maintains a policy $\pi_i(a_i)$, i.e. a distribution over actions and shares. It is initialized with an arbitrary prior distribution. A rational agent wants to optimize its policy such that the expectation of reward is maximized: $\max \mathbb{E}_{a} r_i(a)$, where $a \sim \times_{i \in N} \pi_i(a_i)$. Note that optimization of an individual's policy depends on the policies of all other agents. Also note that policy optimization of self-interested individuals is non-stationary due to changes in others' policies as they optimize for themselves.

After initialization, DOS performs the following steps for a predefined number of iterations.

\begin{enumerate}
	\item Each agent samples a multiset of $n_\mathrm{sample}$ actions from its policy and communicates it to other agents.
	\item A list of joint actions is constructed from the communicated action lists of other agents.
	\item The utility of each joint action is determined according to Equation \ref{eq:dos_util}.
	\item The policy is updated in a way that increases the likelihood of sampling high-utility actions and shares.
\end{enumerate}

After $n_\mathrm{iter}$ iterations, each agent samples an action and a share from its policy, executes the action, and shares reward accordingly. The resulting joint action yields the global result of DOS.

\begin{algorithm}
	\begin{algorithmic}[1]
		\State initialize $\pi_i$ for each agent $i$
		\For {$n_\mathrm{iter}$ iterations}
			\For {each agent $i$}
				\State each agent samples a list of $n_\mathrm{sample}$ actions from $\pi_i$
				\State broadcast sampled actions
			\EndFor
			\For {each agent $i$}
				\State build joint actions $a$
				\State determine utility $u_i(a)$ according to Eq. \ref{eq:dos_util}
				\State update $\pi_i$ to increase the likelihood of high-utility samples
			\EndFor
		\EndFor
		\For {each agent $i$}
			\State execute $a_i$ and share $s_i(a)$ sampled from $\pi_i$
		\EndFor
	\end{algorithmic}
	\caption{Distributed Optimization with Sharing (DOS)}
	\label{alg:dos}
\end{algorithm}

\subsection{Cross-Entropy DOS}

In general, DOS is parametric w.r.t. modeling and updating of policies $\pi_i$. As an example, we instantiate DOS with cross entropy optimization \cite{kroese2013cross}. We label this instantiation CE-DOS.

For CE-DOS, we model a policy $\pi$ as isotropic normal distribution $\mathcal{N}(\mu, \sigma)$. I.e., each parameter of an action is sampled from a normal distribution that is independent from other action parameter distributions. Note that it is also possible to model policies in terms of normal distribution with full covariance, but the simpler and computationally less expensive isotropic representation suffices for our illustrative concerns. As prior CE-DOS requires initial mean $\mu_0$ and standard deviation $\sigma_0$ for a policy (cf. Algorithm \ref{alg:cedos}, line 1). I.e. initial actions before any optimization are sampled as follows.
\begin{align}
\label{eq:ce_dos_prior}
a_i \sim \mathcal{N}(\mu_0, \sigma_0)
\end{align}

Updating a policy (cf. Algorithm \ref{alg:dos}, line 12 - 15) is done by recalculating mean and variance of the normal distribution. We want the update to increase the expected sample utility. For each of $n_\mathrm{iter}$ iterations, we sample $n_\mathrm{sample}$ actions and shares $a_i, s_i \sim \pi_i$ from each agent's policy, and build the corresponding joint actions $a = \times_{i \in N} a_i$ and shares $s = \times_{i \in N} s_i$.

Each agent evaluates sampled actions and shares according to its utility $u_i(a, s)$.
From the set of evaluated samples of each agent, we drop a fraction $\psi \in (0, 1]$ of samples from the set wrt. their utilities. That is, we only keep high utility samples in the set. We then compute mean and variance of the action parameters in the reduced set, and use them to update the policy. A learning rate $\alpha \in (0, 1]$ determines the impact of the new mean and variance on the existing distribution parameters: E.g. let $\mu_t$ and $\sigma_t$ be the mean and standard deviation of a normal distribution modeling a policy at iteration $t$, then
\begin{align*}
\mu_{t + 1} &= (1 - \alpha) \mu_{t} + \alpha \mu_\mathrm{new}\\
\sigma_{t + 1} &= (1 - \alpha) \sigma_{t} + \alpha \sigma_\mathrm{new}
\end{align*}
where $\mu_\mathrm{new}$ and $\sigma_\mathrm{new}$ are mean and standard deviation of the elite samples. We require a lower bound $\sigma_\mathrm{min}$ on the standard deviation of policies in order to maintain a minimum amount of exploration.

The hyperparameters of CE-DOS are thus as follows.
\begin{itemize}
	\item A stochastic game $(X, N, A, p, R)$
	\item Number of iterations $n_\mathrm{iter}$ 
	\item Number of samples $n_\mathrm{sample}$ from the policy at each iteration
	\item Prior mean $\mu_0$ and standard deviation $\sigma_0$ for policies
	\item Lower bound $\sigma_\mathrm{min}$ on the policy standard deviations
	\item Fraction $\psi \in (0, 1]$ of elite samples to keep
	\item Learning rate $\alpha \in (0, 1]$
\end{itemize}

\begin{algorithm}
	\begin{algorithmic}[1]
		\State Intitialize $\pi_i \gets \mathcal{N}(\mu_0, \sigma_0)$ for each agent $i$
		\For {$n_\mathrm{iter}$ iterations}
		\For {each agent $i$}
		\State sample $n_\mathrm{sample}$ actions and shares $a_i, s_i \sim \pi_i$
		\State clip $s_i$ such that $s_i \geq 0$
		\State broadcast sampled actions and shares
		\EndFor
		\For {each agent $i$}
		\State build joint actions $a = \times_{i \in N} a_i$ and shares $s = \times_{i \in N} s_i$
		\State determine utility $u_i(a, s)$ according to Eq. \ref{eq:dos_util}
		\State keep $\psi \cdot n_\mathrm{sample}$ elite samples $a, s$ with highest utility
		\State compute $\mu_\mathrm{new}$ and $\sigma_\mathrm{new}$ from $a_i, s_i$ in the elite samples
		\State $\mu_{t + 1} \gets (1 - \alpha) \mu_{t} + \alpha \mu_\mathrm{new}$
		\State $\sigma_{t + 1} \gets (1 - \alpha) \sigma_{t} + \alpha \sigma_\mathrm{new}$
		\State $\sigma_{t + 1} \gets \max(\sigma_{t + 1}, \sigma_{\min})$
		\State $\pi_i \gets \mathcal{N}(\mu_{t + 1}, \sigma_{t + 1})$
		\EndFor
		\EndFor
		\For {each agent $i$}
		\State $a_i, s_i \sim \pi_i$
		\State execute $a_i$ and share $s_i$
		\EndFor
	\end{algorithmic}
	\caption{Cross Entropy DOS}
	\label{alg:cedos}
\end{algorithm}

\section{Experimental Results and the Sharer's Dilemma}
\label{sec:results}

We experimentally analyzed the effects of sharing in collective adaptive systems of self-interested agents.

\subsection{Domains}

We evaluated the effect of sharing utilities with CE-DOS in two synthetic domains. In these domains, a CAS of self-interested agents has to balance individual and global resource consumption (or production, respectively).

For example, the energy market in the smart grid can be modeled as a CAS of self-interested agents. Each participant has to decide locally how much energy to produce. Each agent wants to maximize its individual payoff by selling energy to consumers in the grid. Therefore, each agent would like to maximize its individual energy production. However, the selling price per unit is typically non-linearly depending on global production. For example, global overproduction is penalized.

There are a number of corresponding real world problems, for example energy production and consumption in the smart grid, traffic routing, passenger distribution to individual ride hailing participants, cargo distribution on transport as a service, routing of packets in networks, distribution of computational load to computers in a cluster, and many more.

We now define two market models (simple and logistic) as domains for evaluating the effects of sharing in CAS of self-interested agents.

\subsubsection{Simple Market}
We model individual and global production, and use their relation for calculating utilities in such a scenario. We set $A_i = \mathbb{R}^1$ as individual agents' action space, $a_i \in A_i$ models the production amount. The sum $\sum_{i \in N} a_i$ models the global production.

We define the reward of each agent as the relation of its own individual resource consumption to the global resource consumption. I.e. the reward correlates to an agents market share. We introduce a slope parameter $\xi$ to control the utility slope of individual and global consumption.
\begin{equation}
\label{eq:domain}
r_i(a) = \dfrac{a_i}{\left( \sum_{j \in N} a_j \right) ^\xi} 
\end{equation}

In this setup, a rational agent would like to increase its own consumption until saturation.
I.e. a monopoly is able to produce cheaper than two small producers, and therefore an inequal production amount unlocks more global reward. If all agents act rationally by maximizing their individual $a_i$, in general the corresponding equilibrium is not equal to the global optimum.



\subsubsection{Logistic Market}

We modeled another market scenario for investigating the effects of sharing in CAS of self-interested agents. As before, each agent has to choose the amount of energy to use for production of a particular good. I.e. $A_i \in [.1, 4]$, as in the simple market domain. Note that this is an arbitrary choice.

Each agent has a logistic production curve $p_i : A_i \rightarrow [0, 1]$ as a function of its invested energy. For example, this models different production machine properties. The logistic curve $p_i$ is given as follows.
\begin{equation}
	p_i(a_i) = \dfrac{1}{1 + e^{- c (a_i - o)}}
\end{equation}
Here, $c \in \mathbb{R}$ defines the steepness of the logistic function, and $o \in \mathbb{R}$ determines the offset on the x-axis.

Global production $\mathit{prod}$ is the sum of individual production $\sum_i p_i(a_i)$.
A price function (i.e. an inverse logistic function) defines the price per produced unit, given global production $\mathit{prod}$.
\begin{equation}
\mathit{price}(\mathit{prod}) = 1 - \dfrac{1}{1 + e^{- c (\mathit{prod} - o)}}
\end{equation}

The reward for an agent is defined as the product of its produced units and the global price.
\begin{equation}
r_i(a) = p_i(a_i) \cdot \mathit{price}(\mathit{prod}) ~ \mathrm{where} ~ \mathit{prod} = \sum_{j \in N} p_j(a_j)
\end{equation}

\begin{figure}
	\centering
	\includegraphics[width=.45\textwidth]{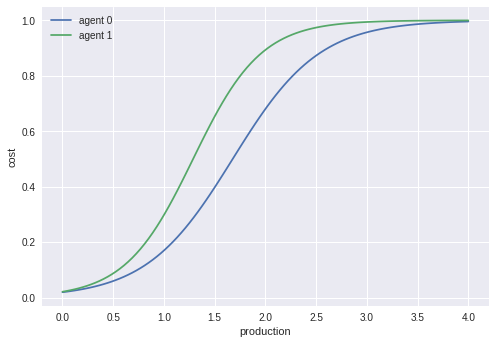}
	\includegraphics[width=.45\textwidth]{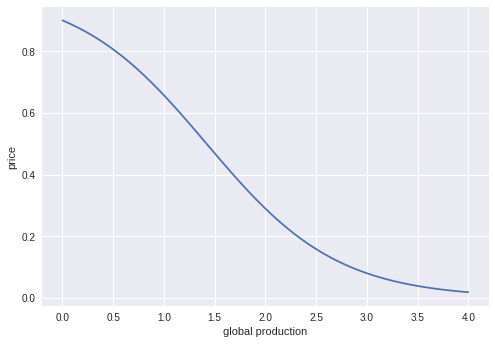}
	\caption{Example production functions (left) and global price function (right) in the logistic market domain.}
	\label{fig:market_domain}
\end{figure}

Figure \ref{fig:market_domain} shows an example of production and price functions in the logistic market domain.

\subsection{Setup}

For our experiments, we used the following setup of CE-DOS.\footnote{We plan to publish our code upon publication.}
\begin{itemize}
\item We consider a stochastic game with $n$ agents, that is $N = \{1, ..., n\}$.
\item We set $n = 10, n = 50$ and $n = 100$ in our experiments.
\item Individual action spaces were set as $A_i = [.1, 4]$.
\item We define the individual reward functions as given by Equation \ref{eq:domain}.
\item We set the number of iterations $n_\mathrm{iter}$ for CE-DOS to 100.
\item We draw $n_\mathrm{sample} = 100$ samples from the policy per iteration for each agent.
\item Prior mean $\mu_0$ and standard deviation $\sigma_0$ were set to 0 and 1, respectively.
\item We set the fraction of elite samples $\psi = 0.25$.
\item We set the learning rate $\alpha = 0.5$.
\item We set the minimal policy standard deviation $\sigma_{\min} = 0.2$.
\end{itemize}

We sampled domain parameters uniformely from the following intervals.
\begin{itemize}
	\item We sampled the slope parameter $\xi$ from $[2, 4]$ in the simple market domain.
	\item We sampled logistic steepness $c$ and offset $o$ from $[1, 3]$ for all production and cost functions in our experiments with the logistic market domain.
\end{itemize} 

We varied the number of sharing agents to measure the effect of defecting (i.e. non-sharing) agents that participate in the stochastic game together with sharing individuals.

Note that for the results we report here, we clipped the sharing values such that agents are only able to share up to their current reward, i.e. $s_i \leq r_i(a)$ for a given $a \in A$. In general, other setups with unbound sharing are possible as well.

\todo[inline]{correlation of effect and number of involved agents}
\todo[inline]{illustrate optimization trajectories in utility landscape}

\subsection{Effect of Sharing on Global Reward}

\begin{figure}
	\centering
	\includegraphics[width=.49\textwidth]{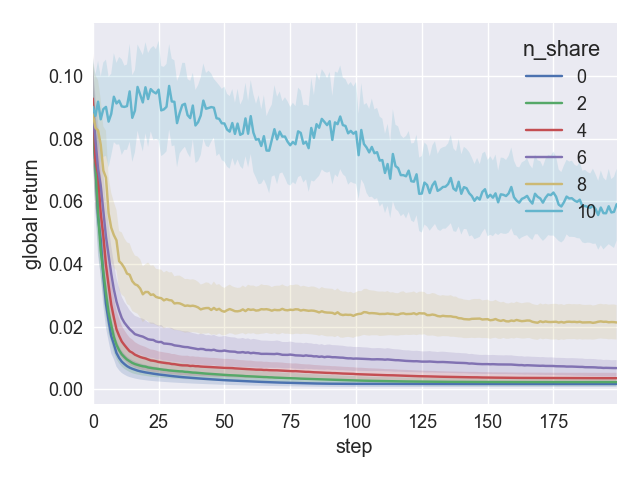}
	\includegraphics[width=.49\textwidth]{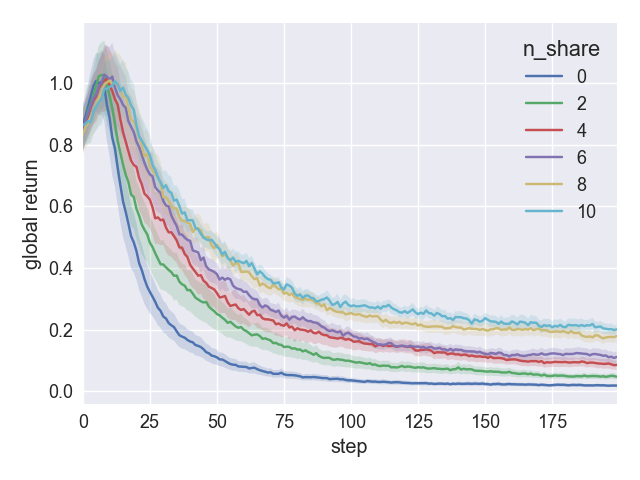}
	\includegraphics[width=.49\textwidth]{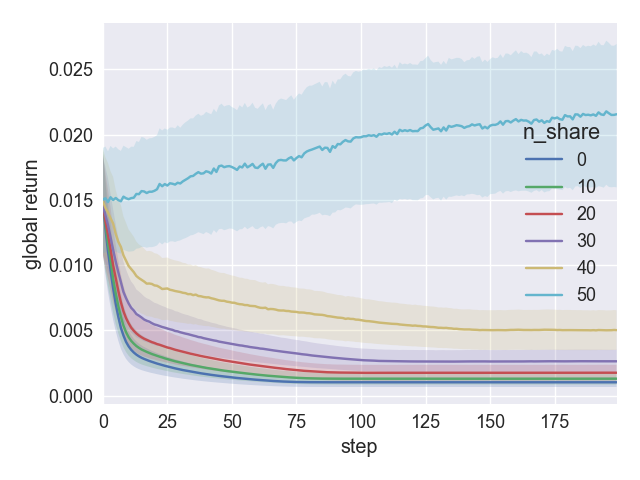}
	\includegraphics[width=.49\textwidth]{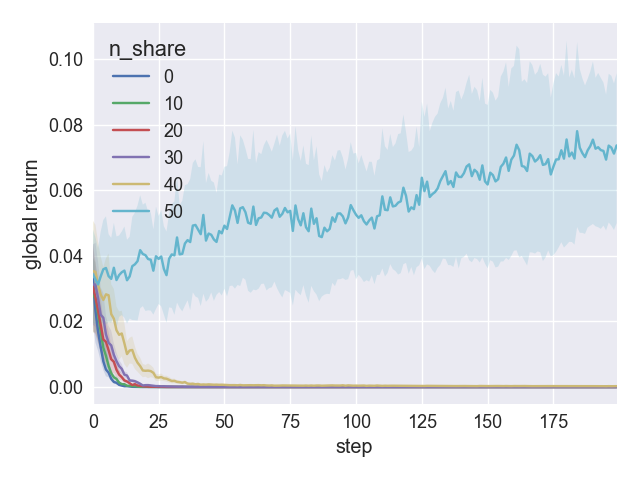}
	\includegraphics[width=.49\textwidth]{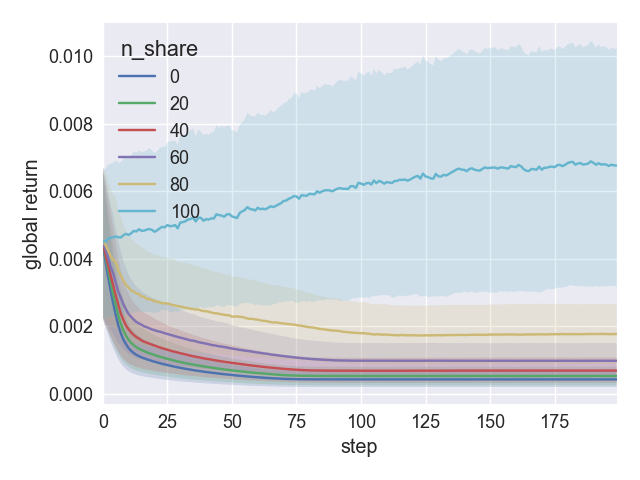}
	\includegraphics[width=.49\textwidth]{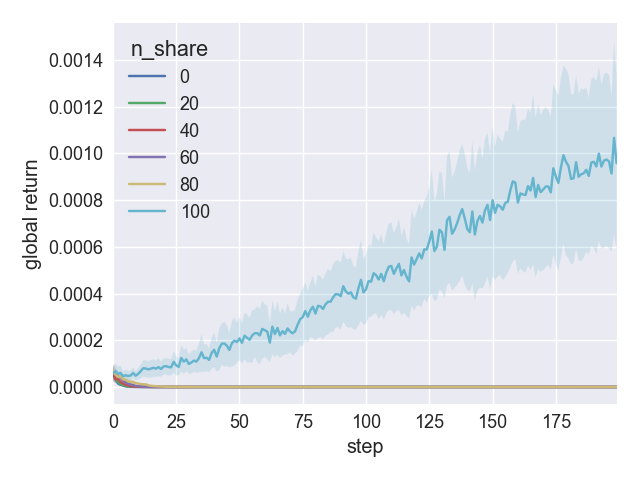}
	\caption{Global utility gathered for varying numbers of sharing agents in the simple market (left column) and logistic market (right column) domains. 10 agents (top row), 50 agents (center row) and 100 agents (bottom row) in total. Solid line shows empirical mean of 10 experimental runs, shaded areas show .95 confidence intervals. Best viewed on screen in color.}
	\label{fig:values}
\end{figure}

Figure \ref{fig:values} shows the mean global utility gathered for varying numbers of sharing agents. We can observe that the fraction of sharing agents correlates with global utility. We also see that the effect of sharing increases with the number of participating agents.

\begin{figure}
	\centering
	\includegraphics[width=.49\textwidth]{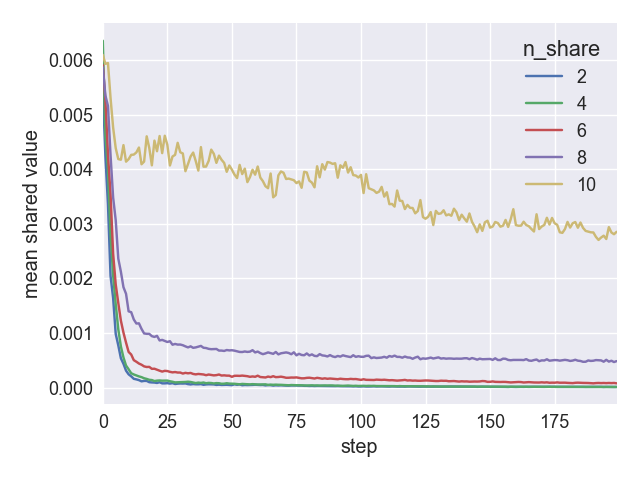}
	\includegraphics[width=.49\textwidth]{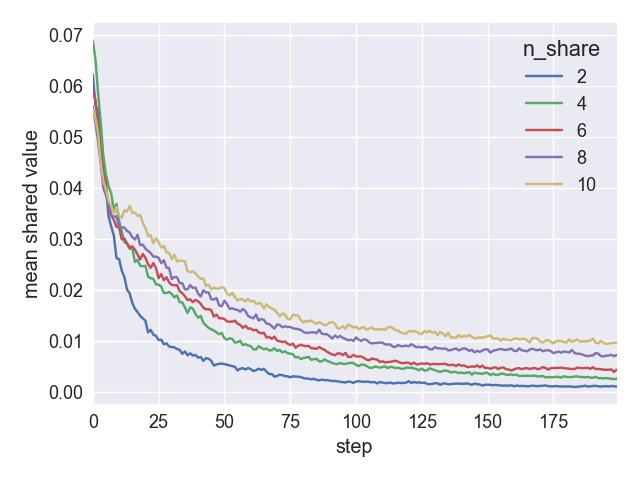}
	\includegraphics[width=.49\textwidth]{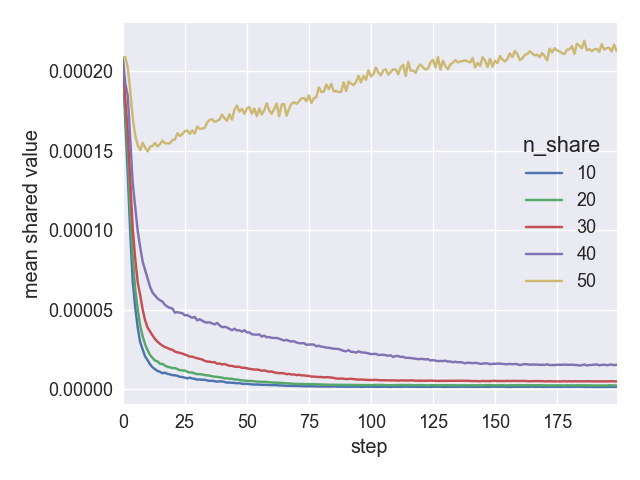}
	\includegraphics[width=.49\textwidth]{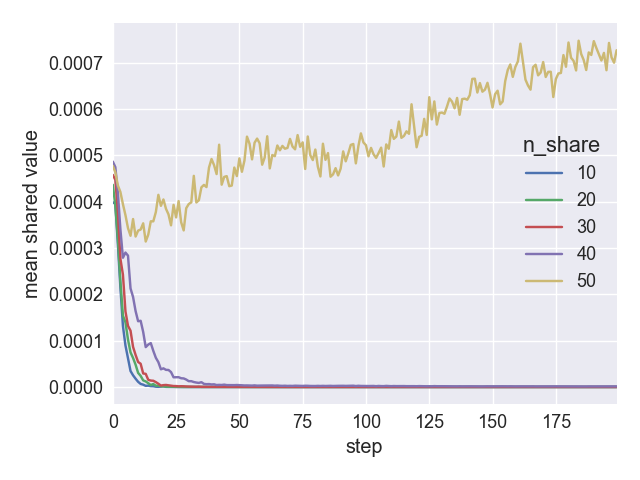}
	\includegraphics[width=.49\textwidth]{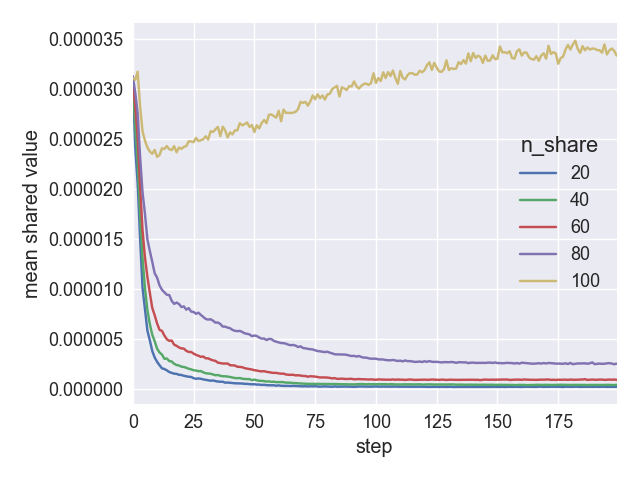}
	\includegraphics[width=.49\textwidth]{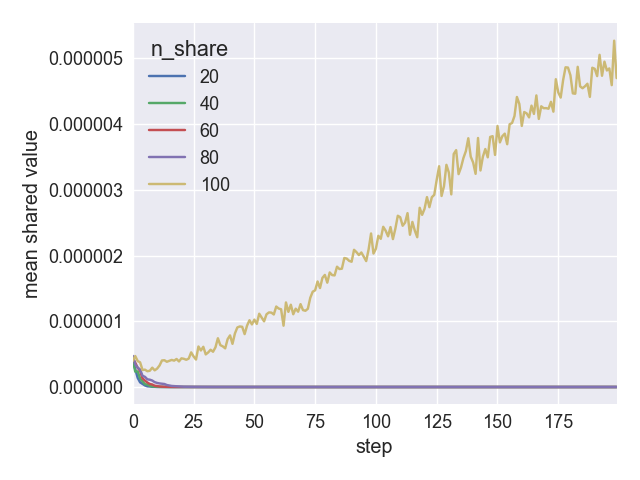}
	\caption{Mean individual shares for varying numbers of sharing agents in the simple market (left column) and logistic market (right column) domains. 10 agents (top row), 50 agents (center row) and 100 agents (bottom row) in total. Solid line shows empirical mean of 10 experimental runs, shaded areas show .95 confidence intervals. Best viewed on screen in color.}
	\label{fig:shares}
\end{figure}

Figure \ref{fig:shares} shows the mean individual shared value for the corresponding experimental setups. We can see that the amount of shared value correlates with global reward. I.e. the more value shared, the higher the global reward. We also see that the number of participating agents correlates with the effect of sharing.

\subsection{Sharer's Dilemma}

\begin{figure}
	\centering
	\includegraphics[width=.49\textwidth]{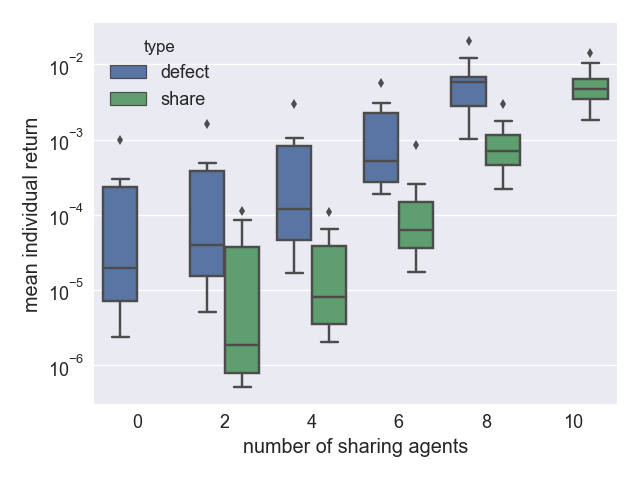}
	\includegraphics[width=.49\textwidth]{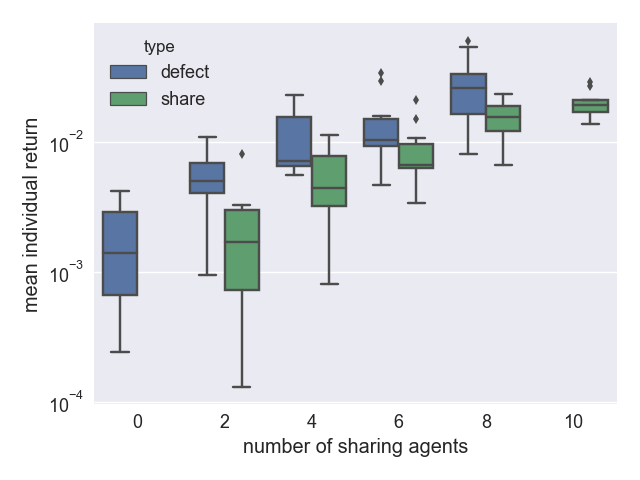}
	\includegraphics[width=.49\textwidth]{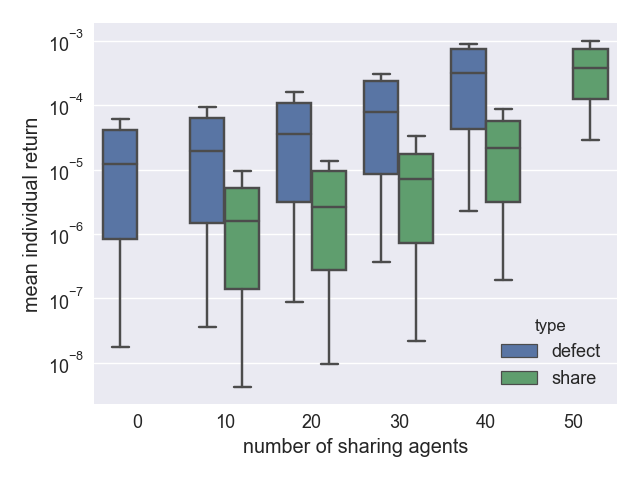}
	\includegraphics[width=.49\textwidth]{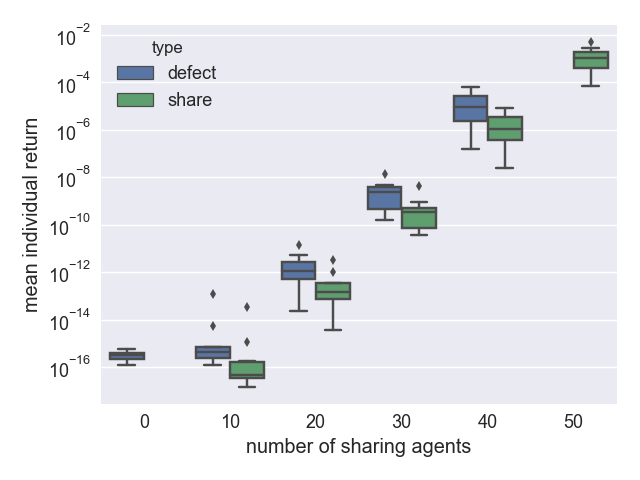}
	\includegraphics[width=.49\textwidth]{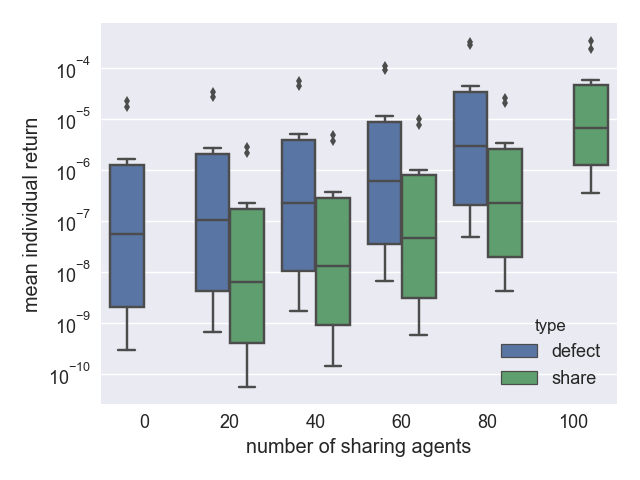}
	\includegraphics[width=.49\textwidth]{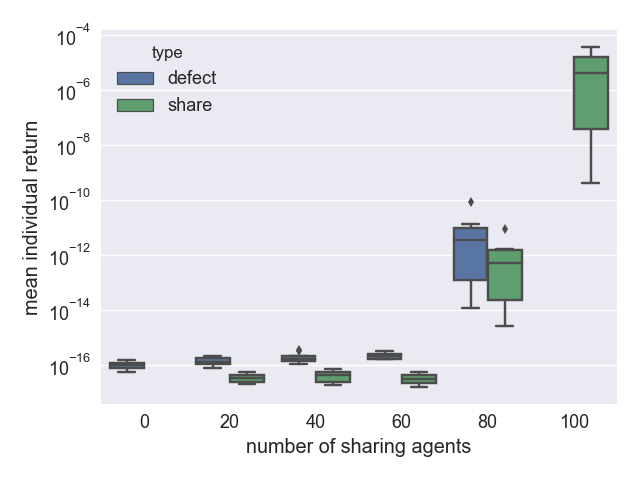}
	\caption{Schelling diagrams showing mean individual utility for defectors and sharers, for varying numbers of sharing agents in the simple market (left column) and logistic market (right column) domains. Note the log scale on the y-axis. 10 agents (top row), 50 agents (center row) and 100 agents (bottom row) in total. 10 experimental runs. Best viewed on screen in color.}
	\label{fig:schelling}
\end{figure}

Figure \ref{fig:schelling} shows the Schelling diagrams for the corresponding experiments. A Schelling diagram compares the mean individual utility of sharers and defectors based on the global number of sharing agents \cite{schelling1973hockey}. We can see that agents that choose to defect gather more individual utility than the sharing ones.

The shape of the Schelling diagrams in Figure \ref{fig:schelling} shows that sharing in collective adaptive systems with self-interested agents yields a dilemma in our experimental setups.

\begin{center}
	Should an individual agent share or defect?
\end{center}

There is no rational answer to this question for an individual self-interested agent. If the agent chooses to share, it may be exploited by other agents that are defecting. However, if the agent chooses to defect, it may hurt its individual return by doing so in comparison to having chosen to share.

Note that the amount of sharing is a free parameter to be optimized by DOS. This means that all behavior we observe in our experiments is emergent. The combination of available resources, interdependency of agents' actions and the ability to share lets agents decide to share with others based on their intrinsic motivation.

Our results illustrate a potential reason for emergence of cooperation and inequity aversion in CAS of only self-interested agents. They also give an explanation to the existence of punishment of individuals that exploit societal cooperation at the cost of sharing individuals' and global reward.

\section{Conclusion}
\label{sec:conclusion}

We summarize the ideas in this paper, discuss limitations and implications of our results, and outline venues for further research.

\subsection{Summary}

In collective adaptive systems (CAS), adaptation can be implemented by optimization wrt. utility. Agents in a CAS may be self-interested, while their utilities may depend on other agents' choices. Independent optimization of each agent's utility may yield poor individual and global payoff due to locally interfering individual preferences in the course of optimization. Joint optimization may scale poorly, and is impossible if agents do not want to expose their preferences due to privacy or security issues.

In this paper, we studied distributed optimization with sharing for mitigating this issue. Sharing utility with others may incentivize individuals to consider choices that are locally suboptimal but increase global reward. To illustrate our ideas, we proposed a utility sharing variant of distributed cross entropy optimization. Empirical results show that utility sharing increases expected individual and global payoff in comparison to optimization without utility sharing.

We also investigated the effect of defectors participating in a CAS of sharing, self-interested agents. We observed that defection increases the mean expected individual payoff at the expense of sharing individuals' payoff. We empirically showed that the choice between defection and sharing yields a fundamental dilemma for self-interested agents in a CAS.

\subsection{Limitations}

A central limitation of CE-DOS is its state- and memoryless optimization.
In our formulation of utility sharing self-interested agents optimize an individual action and share that maximizes their utility.
However, our formulation does not account for learning decision policies based on a current state and other learning agents. In this case, the utility of each agent would also depend on concrete states, transition dynamics and potentially also on models agents learn about other participants \cite{foerster2017learning,rabinowitz2018machine}.

As there is no temporal component to the optimization problems that we studied in this paper, it is also not possible to study the effect of gathering wealth in our current setup. We think that the dynamics of sharing in temporally extended decision problems may differ from the ones in stateless optimization. For example, corresponding observations have been made for game theoretic dillemas, where optimal strategies change when repeating a game (in contrast to the optimal strategy when the game is only played once) \cite{sandholm1996multiagent}. Similar research has been conducted in the field of reinforcement learning, however not accounting for utility sharing so far \cite{leibo2017multi}.

We also want to point out that exposing shares eventually provides ground for attack for malicious agents \cite{brundage2018malicious}. Albeit indirectly, exposed shares carry information about individual utility landscapes, allowing attackers to potentially gather sensitive information about agents' internal motivations. Agents in critical application domains should consider this weakness when opting to share.


\subsection{Future Work}

In future work, we would like to transfer our approach to temporally extended domains and model sharing in CAS with multi-agent reinforcement learning. Hopefully, this would enable studying sharing and the Sharer's Dilemma in more complex domains.

We also think that there are many interesting options for realizing sharing besides equal distribution as formulated in Eq. \ref{eq:dos_util}. For example, our formulation does not allow for bilateral shares or formation of coalitions.
Also, we would be interested to study the effect of wealth on emergent cooperation and defection.
Another interesting line would be to investigate the effects of punishment in CAS of self-interested agents.

As an application domain, it would be interesting to exploit the duality of planning and verification. For example, agents utility could model individual goal satisfaction probability. Sharing could be used to increase individual and global goal satisfaction probability in CAS.


\bibliographystyle{llncs2e/splncs}
\bibliography{references}

\end{document}